\documentclass[12pt]{article}
\usepackage{amscd}
\usepackage[]{hyperref}
\usepackage{amsmath,amssymb,latexsym,cite}
\usepackage{bm}
\input xy
\xyoption{all}
\usepackage{bbold}
\usepackage{mathtools}

\newcommand{\be}{\begin{equation}}
\newcommand{\ee}{\end{equation}}
\newcommand{\bea}{\begin{eqnarray}}
\newcommand{\eea}{\end{eqnarray}}

\newcommand{\no}{\nonumber}

\def\l:{\mathopen{:}\,}
\def\r:{\,\mathclose{:}}

\newcommand{\cO}{{\mathcal{O}}}
\newcommand{\cS}{{\mathcal{S}}}

\newcommand{\cQ}{{\mathcal{Q}}}
\newcommand{\cN}{{\mathcal{N}}}

\DeclareMathOperator{\Det}{Det}

\begin{document}

\vspace*{0.5in}

\begin{center}

{\Large \bf Quantum cohomology of symplectic flag manifolds}

\vspace{0.2in}

Jirui Guo\footnote{{\tt jrguo@mail.tsinghua.edu.cn}}, Hao Zou\footnote{{\tt hzou@vt.edu}}

\vspace*{0.2in}

\itshape{$^1$ Yau Mathematical Sciences Center, Tsinghua University, Beijing 100084, China \\
$^2$ Department of Physics, Virginia Tech, 850 West Campus Dr. Blacksburg, VA 24061}

$\,$

\end{center}

\hfill

\begin{abstract}
\noindent We compute the quantum cohomology of symplectic flag manifolds. Symplectic flag manifolds can be described by non-abelian GLSMs with superpotential. Although the ring relations cannot be directly read off from the equations of motion on the Coulomb branch due to complication introduced by the non-abelian gauge symmetry, it can be shown that they can be extracted from the localization formula in a gauge-invariant form. Our result is general for all symplectic flag manifolds, which reduces to previously established results on symplectic Grassmannians and complete symplectic flag manifolds derived by other means. We also explain why a $(0,2)$ deformation of the GLSM does not give rise to a deformation of the quantum cohomology.
\end{abstract}

\newpage

\tableofcontents

\newpage

\section{Introduction}

Since its appearance in the context of topological quantum field theories \cite{W1, W2}, quantum cohomology has drawn tremendous interest among physicists and mathematicians (see, e.g. \cite{Cecotti:1991vb,Vafa:1991uz,Bat93,Witten:1993xi,Morrison:1994fr,Kontsevich:1994qz,Ruan:95,Kim:1994ny,Givental:95}). It is a variant of the ordinary cohomology ring of the target space, and encodes enumerative data known as Gromov-Witten invariants (see, e.g.\cite{Kontsevich:1994qz,Ruan:95,Fulton:96} and \cite[Chapter 7]{CK}). Physically, the structure constants of the quantum cohomology ring are determined by three-point functions of the corresponding A-twisted nonlinear sigma model (NLSM), from which Yukawa couplings of the effective spacetime theory can be read off. If the NLSM admits a UV completion, namely a gauged linear sigma model (GLSM) \cite{W3}, the computation of correlation functions can be greatly simplified by supersymmetric localization \cite{BZ, CCP,CGJS} in many cases.

The ring structures of quantum cohomology are known for many families of spaces (see, e.g. \cite{Intriligator:1991,Bat93,Witten:1993xi,Kim:1994ny,Givental:95,Astashkevich:1993ks,Buch:01,Buch:03,B}). In this work, we derive the ring structure for symplectic flag manifolds, which can be described by non-abelian GLSMs \cite{GSZ}. This problem has been studied before in two special cases, namely the symplectic Grassmannians \cite{KT, BKT} and the complete symplectic flag manifolds \cite{K}. Our computation generalizes these results and gives rise to a representation for the quantum cohomology ring of an arbitrary symplectic flag manifold.

For (symplectic) flag manifolds, the mirror map is trivial, or mathematically speaking the $I$-function equals the $J$-function. Consequently, the quantum cohomology ring is fully determined by the Jacobi ring of the mirror superpotential\footnote{We thank Wei Gu for pointing this out.}. Therefore the equations of motion on the Coulomb branch of the GLSM are sufficient to determine all the ring relations. %The symplectic flag manifolds we are considering in this paper are within these situations. 
Whereas in abelian GLSMs such as those for toric varieties, these equations of motion are precisely the quantum cohomology relations, it is not a trivial task to extract the gauge-invariant relations from them for non-abelian theories. In this paper we adopt the method used in \cite{G}. Generally speaking, this method consists of two steps. First, one uses the classical limit of the localization formula \cite{CCP, CGJS} on the Coulomb branch to select a set of classical ring relations. Second, one computes the quantum corrections to these relations from the equations of motion on the Coulomb branch. (We review the computation for the quantum sheaf cohomology of ordinary flag manifolds in the appendix.) The first step is an efficient way of selecting a complete set of gauge-invariant relations in the classical limit.

Our main result is as follows. For the symplectic flag manifold $SF(k_1,\cdots,k_n,2N)$, i.e. flags of isotropic subspaces with dimensions $k_1,\cdots,k_n$ inside a $2N$-dimensional symplectic complex vector space, there exist a set of generators labeled by $x^{(s)}_{i_s}$ with $s=1,\cdots,n+1$ and $i_s \in \mathbb{Z}_{\geq 1}$. The degree of $x^{(s)}_{i_s}$ is $i_s$. If we denote by $\mathcal{A}_n$ the polynomial ring generated by the variables $x^{(s)}_{i_s}$, then the quantum cohomology of $SF(k_1,\cdots,k_n,2N)$ is the quotient ring
\be \label{eq:qcsf}
\mathcal{A}_n /(I+R_1+R_2),
\ee
where $I$ encodes the relations
\[
\sum_{i_1+i_2+\cdots+i_{n+1}=r}x^{(1)}_{i_1} x^{(2)}_{i_2} \cdots x^{(n+1)}_{i_{n+1}} = 0,\quad r>0.
\]
$R_1$ is generated by the relations
\[
x^{(1)}_r = 0,~r>k_1,
\]
and
\[
x^{(s)}_r =
\left\{ \begin{array}{ll}
0, & \mathrm{if}~ k_s-k_{s-1} < r < k_s-k_{s-2}, \\
(-)^{k_{s-1}+k_{s-2}-1} q_{s-1} y^{(s-1)}_{r-k_s+k_{s-2}}, & \mathrm{if}~ r \geq k_s-k_{s-2}.
\end{array} \right. ~s=2,\cdots,n.
\]
$R_2$ is generated by
\begin{equation}\label{R2}
\sum_{i+j=2r}(-1)^i x^{(n+1)}_i x^{(n+1)}_j =
\left\{ \begin{array}{l}
0, \quad \quad \mathrm{if}~ 2(N-k_n+1) \leq 2r < 2N-k_n-k_{n-1}+1, \\
(-)^{k_{n}+k_{n-1}-1} q_n y^{(n)}_{2r-(2N-k_n-k_{n-1}+1)}, \quad \mathrm{if}~ 2r \geq 2N-k_n-k_{n-1}+1.
\end{array} \right.
\end{equation}
If we denote by $\mathcal{S}_i$ the $i$-th tautological bundle over $SF(k_1,\cdots,k_n,2N)$, then  $x^{(s)}_r$ can be identified with the $r$-th Chern class of $\mathcal{S}_{s}/\mathcal{S}_{s-1}$ with $\mathcal{S}_0 = 0$ and $\mathcal{S}_{n+1} = \mathcal{O}^{\oplus(2N)}$. The left hand side of \eqref{R2} can be interpreted as the $2r$-th Chern class of $\mathcal{Q}_n/\mathcal{S}_n^\vee$, where $\mathcal{Q}_s = \mathcal{O}^{\oplus(2N)}/\mathcal{S}_s$. In the expressions above, the parameters $q_1,q_2,\cdots,q_n$ parameterize the K\"ahler moduli.

One may as well turn on (0,2) deformations and compute the quantum sheaf cohomology using the same method. (See e.g. \cite{KS, ADE, MM1, MM2, DGKS, GLS} for discussions on (0,2) deformation and quantum sheaf cohomology.) But as we will show in section \ref{sec:02}, the supersymmetric constraint $E^a J_a = 0$ of the GLSM excludes all nontrivial deformation of the tangent bundle. The reason why ordinary flag manifold admits nontrivial (0,2) deformation is due to the absence of $J$-terms in the corresponding GLSM \cite{DS}.

This paper is organized as follows. In section \ref{sec:computation}, we perform the computation and derive the ring structure of the quantum cohomology of general symplectic flag manifolds. In section \ref{sec:reduction}, we show that our result can be reduced to existing mathematical results in the case of symplectic Grassmannians and complete symplectic flag manifolds. In section \ref{sec:02}, we show that there are no nontrivial (0,2) deformations of the A/2-twisted GLSM for symplectic flag manifolds. We review the (0,2) deformation and the quantum sheaf cohomology of ordinary flag manifolds in the appendix for comparison.

\section{Computation on the Coulomb branch}\label{sec:computation}

Physically, quantum cohomology of a K\"ahler manifold $X$ is defined by the OPE of the A-twisted version of the two-dimensional $\cN =2$ nonlinear sigma model with target space $X$. Due to the topological nature, if this NLSM can be realized as a low energy effective theory of some GLSM in the large radius limit, then the quantum product can also be computed by the A-model OPE of the GLSM. A crucial step in this computation is to identify the GLSM operators with the cohomology classes of $X$. This can usually be done on the Coulomb branch, where supersymmetric localization is then employed to compute the correlation functions.

Now we focus on the situation in which the target space is a symplectic flag manifold. For every sequence of integers $(k_1,k_2,\cdots,k_n)$ with $0 < k_1 < k_2 < \cdots < k_n < N$, the symplectic flag manifold $SF(k_1,k_2,\cdots,k_n,2N)$ is defined to be the set of sequences of subspaces $V_1 \subset V_2 \subset \cdots \subset V_n$ in $\mathbb{C}^{2N}$ with dimensions $k_1,k_2,\cdots,k_n$ respectively, such that $V_n$ is isotropic with respect to a symplectic form on $\mathbb{C}^{2N}$. There is a flag of tautological subbundles
\[
	0=\mathcal{S}_0 \hookrightarrow \mathcal{S}_1 \hookrightarrow \mathcal{S}_2 \hookrightarrow \cdots \hookrightarrow \mathcal{S}_n \hookrightarrow \mathcal{S}_{n+1}=\mathcal{O}^{\oplus (2N)},
\]
where the fibers of these bundles at any point of the flag manifold form the flag corresponding to that point, so $\mathcal{S}_i$ has rank $k_i$. There are also tautological quotient bundles defined by the short exact sequences
\[
0 \rightarrow \cS_m \rightarrow \mathcal{O}^{\oplus(2N)} \rightarrow \cQ_m \rightarrow 0
\]
for $m = 1,\cdots,n$.

Now we give a description of the classical cohomology ring of $SF(k_1,k_2,\cdots,k_n,2N)$ which is suitable for applying our approach to compute the quantum corrections. Let $x^{(m)}_i$ be the $i$-th Chern class of the quotient bundle $\mathcal{S}_{m}/\mathcal{S}_{m-1}$, $m=1,\cdots,n+1$. Define
\[
[x^{(s)}] \:=\: \sum_{i \geq 0} x_i^{(s)}
\]
with the convention $x_0^{(s)} = 1$,
so $[x^{(s)}]$ is the total Chern class of $\mathcal{S}_{m}/\mathcal{S}_{m-1}$.
The cohomology ring of $SF(k_1,k_2,\cdots,k_n,2N)$ can be written as
\be\label{AIR1R2}
\mathcal{A}/( I+R_1+R_2 ),
\ee
where $I$ encodes the fact that $\mathcal{S}_{n+1}$ is trivial, or more explicitly
	\be
		c\left(\cS_1\right) \cdot c\left( \cS_2/\cS_1 \right) \cdots c\left( \cS_i/\cS_{i-1} \right) \cdots c\left( \cO^{2N}/\cS_n \right) \:=\: [x^{(1)}][x^{(2)}]\cdots[x^{(n+1)}] \:=\: 1
	\ee
due to the Whitney sum formula. Now we see the ideal $I$ is independent of quantum deformation as it is the constraint that $\{x^{(s)}_r\}$'s should satisfy. The quantum deformations will only affect $R_1$ and $R_2$.

$R_1$ is generated by $x^{(m)}_r$ for $m=1,\cdots,n$ and $r > k_m-k_{m-1}$, 
which is a consequence of the fact that $\mathcal{S}_m/\mathcal{S}_{m-1}$ has rank $k_m-k_{m-1}$.

The symplectic form in the definition of the symplectic flag manifold gives a pairing
	\[
		\mathcal{S}_n \otimes \mathcal{Q}_n \longrightarrow \mathcal{O}.
	\]
This pairing induces an injection $\iota: \mathcal{S}_n^\vee \hookrightarrow \mathcal{Q}_n$, so the quotient bundle $\mathcal{Q}_n/\mathcal{S}_n^\vee$ is well-defined and its rank is\footnote{In the maximal case, i.e. $k_n = N$, the injection $\iota$ becomes an isomorphism, $\mathcal{S}_n^\vee \cong \mathcal{Q}_n$, and this will impose stronger constraints $c_i\left(\cS\right)\:=\: (-)^i c_i \left( \cQ \right)$.
	%We will make more comments on this in section \ref{}.
}
	\[
		\mathrm{rank}\left(\mathcal{Q}_n/\mathcal{S}_n^\vee\right) \:=\: 2\left( N - k_n\right).
	\]
	As a result, $R_2$ is generated by
\be\label{R2one}
c_r(\mathcal{Q}_n/\mathcal{S}_n^\vee) \:=\: \sum_{i+j=r}(-)^i x^{(n+1)}_i x^{(n+1)}_j,~r>2\left( N - k_n\right),
\ee
or
	\be\label{R2two}
		c_r\left(\mathcal{Q}_n\right) - c_r\left(\mathcal{S}^\vee_n\right) \:=\: x^{(n+1)}_r -  [x^{(1)}]^\vee \cdots [x^{(n)}]^\vee |_{r},~r>2\left( N - k_n\right).
	\ee
In the expressions above,
\be
	[x^{(s)}]^\vee \:=\: \sum_{i=0}^{\infty}(-)^i x^{(s)}_{i},
\ee
and $[x^{(1)}]^\vee \cdots [x^{(n)}]^\vee |_{r}$ stands for the sum of terms with degree-$r$ in $[x^{(1)}]^\vee \cdots [x^{(n)}]^\vee$. One can show inductively that the ideals generated by \eqref{R2one} and \eqref{R2two} are identical, which is what we call $R_2$. Moreover, from the relations encoded in $R_1$, it is easy to see that vanishing of \eqref{R2two} implies $x^{(n+1)}_r = 0$ for $r > 2N-k_n$, so the constraint on the rank of $\cQ_n$ is automatically satisfied.

Now let us turn to the physics description. The physics setup for $SF(k_1,k_2,\dots,k_n,2N)$ is a $\cN=(2,2)$ GLSM with gauge group \cite[Section~2.6]{GSZ}
\be
	U(k_1) \times U(k_2) \times \cdots \times U(k_n).
\ee
The matter content consists of
\begin{itemize}
 	\item Chiral multiplets in the bifundamental representations $({\bf k_1}, \overline{\bf k_2} )$, $({\bf k_2}, \overline{\bf k_3})$, $\cdots$, $({\bf k_{n-1}}, \overline{\bf k_n} )$, denoted as $\Phi_{(1,2)}, \Phi_{(2,3)},\cdots, \Phi_{(n-1,n)}$ respectively;
 	\item $2N$ chiral superfields ${\Phi_{(n,n+1)}}^a_i$, $a=1,\cdots,k_n$, $i=1,\cdots,2N$, in the fundamental representation ${\bf k_n}$ of $U(k_n)$;
 	\item One chiral superfield $P_{ab}$ in the anti-symmetric representation $\wedge^2\overline{\bf k_n}$ of $U(k_n)$.
 \end{itemize} 
There is a superpotential
\be
\label{eq:superpotential}
	W\:=\: \sum_{i=1}^N P_{ab} {\Phi_{(n,n+1)}}^a_i {\Phi_{(n,n+1)}}^b_{-i}.
\ee
As reasoned in \cite[Section~2.6]{GSZ}, the geometric phase of this gauge theory realizes the symplectic flag manifold $SF(k_1,k_2,\dots,k_n,2N)$. In order to compute its quantum cohomology, we now focus on the Coulomb branch. On the Coulomb branch, the adjoint valued field strength $\sigma^{(a)}$ of $U(k_a)$ decomposes into the diagonal elements $\{\sigma_1^{(k_a)},\dots,\sigma_{k_a}^{(k_a)}\}$, for $a=1,\dots,n$. At a generic point of the Coulomb branch, the gauge group is a semidirect product of $U(1)$ factors and the Weyl subgroup $\prod_{a=1}^n\mathrm{S}_{k_a}$, where $\mathrm{S}_{k_a}$ is the permutation group of $k_a$ elements that permute $\{\sigma_1^{(k_a)},\dots,\sigma_{k_a}^{(k_a)}\}$. All the generators and relations of the cohomology ring must be invariant under the action of the Weyl group. 

%Let $x^{(m)}_i$ be the $i$-th Chern class of the quotient bundle $\mathcal{S}_{m}/\mathcal{S}_{m-1}$. 

Because $\sigma^{(m)}_a$ is interpreted as the $a$-th Chern root of $\mathcal{S}^\vee_m$, we have
\be\label{xx}
	x^{(m)}_r \:=\: \sum_{i+j=r} h_{i}(\sigma^{(m-1)})(-1)^j e_{j}(\sigma^{(m)}),
\ee
where $e_j(\sigma^{(m-1)})$ is the $j$-th elementary symmetric polynomial of $\sigma^{(m-1)} = \{ \sigma^{(m-1)}_{1},\dots,\sigma^{(m-1)}_{k_{m-1}}\}$ and $h_i(\sigma^{(m)})$ is the $i$-th complete symmetric polynomial of $\sigma^{(m)} = \{ \sigma^{(m)}_{1},\dots,\sigma^{(m)}_{k_{m}}\}$. In particular, for the two special cases $m=1$ and $m=n+1$,
\be
	x^{(1)}_r\:=\: (-)^r e_{r}(\sigma^{(1)}),\quad x^{(n+1)}_r\:=\: h_{r}(\sigma^{(n)}).
\ee
For later convenience, we also introduce variables $y^{(s)}_r$ such that
\be
	[x^{(s)}][y^{(s)}]\:=\:1.
\ee
Then, according to equation~\eqref{xx}, $y^{(s)}_r$ can be written in terms of $\sigma$'s as
\be\label{y}
	y^{(s)}_r \:=\: \sum_{i+j=r} (-1)^i e_{i}(\sigma^{(m-1)}) h_{j}(\sigma^{(m)}).
\ee

The equations of motion (or the chiral ring relations) for the $\sigma^{(a)}_i$ are determined by the one-loop corrected twisted superpotential, which read
\begin{equation}\label{EOM}
\begin{split}
	&\prod_{a=1}^{k_2} m^{(1)}_{\alpha a} \:=\: (-)^{k_1-1} q_1,~ \alpha=1,\cdots,k_1, \\
	&\prod_{b=1}^{k_{s+1}} m^{(s)}_{\alpha b} \:=\: (-)^{k_s-1}q_s \prod_{a=1}^{k_{s-1}} m^{(s-1)}_{a \alpha},~ \alpha=1,\cdots,k_s,~s=2,\cdots,n-1, \\
	&\det (M_\alpha) \:=\: (-)^{k_n-1}q_n \prod_{a=1}^{k_{n-1}} m^{(n-1)}_{a \alpha} \prod_{b=1,b\neq \alpha}^{k_n} \tilde{m}_{\alpha b},~\alpha=1,\cdots,k_n,
\end{split}
\end{equation}
where the matrices $m^{(a)}$, $\tilde{m}$ and $M_{\alpha}$ are defined as follows
\begin{equation}\label{Mblock}
\begin{split}
	m^{(s)}_{a b} &= \sigma^{(s)}_a-\sigma^{(s+1)}_b,\\
	&\quad 	s=1,\cdots, n-1,~~a=1,\cdots,k_{s},~~b=1,\cdots,k_{s+1}, \\
	M_a &= \sigma^{(n)}_a \mathbb{1}_{2N\times 2N} , \\
	\tilde{m}_{ab} &= -\sigma^{(n)}_a-\sigma^{(n)}_b, a \neq b.
\end{split}
\end{equation}
One immediate consequence of equation~(\ref{EOM}) is that the degree of $q_s$ is $k_{s+1}-k_{s-1}$ and, in particular, $\deg q_1 = k_2$ and $\deg q_n = 2N-k_{n-1} - k_n + 1$. 

The main goal of this section is to utilize equation~(\ref{EOM}) to obtain homogeneous gauge-invariant relations, which serve as the quantum cohomology ring relations we are looking for. To this end, we follow the similar strategy used in \cite{G}, motivated by the supersymmetric localization formula \cite{CCP,CGJS}.

\subsection{Algorithm for computing the generating relations}

In general, a quantum cohomology ring can be represented as a quotient
\be\label{quo}
\mathcal{A}/( I+R ),
\ee
where $\mathcal{A}$ is a finitely generated polynomial ring. The ideals $I$ and $R$ encode generating relations. $R$ depends on K\"ahler moduli while $I$ does not. The generators of $\mathcal{A}$ and $I$ can be read off from the identification between cohomology classes and operators on the Coulomb branch. Now we describe the algorithm for computing the ring relations encoded in $R$. We specialize the discussion to symplectic flag manifolds but this algorithm is applicable in many other cases \cite{G} when appropriately adjusted.

A generating relation $\cO_R$ of the quantum cohomology satisfies
\[
	\left< \cO_R \cO\right> \: = \: 0,
\]
for any operator $\cO$, where $\langle \cdot \rangle$ stands for A-model correlation function. On the Coulomb branch, this correlation function can be calculated by supersymmetric localization, while $\cO_R$ and $\cO$ are both polynomials in the $\sigma$'s.

Classically, from the localization formula \cite[Eq.~(4.68)]{CCP}, a generating relation $\mathcal{O}_R$ in the cohomology ring should satisfy (notice that $P_{ab}$ has $R$-charge 2)
\be\label{localization}
\mathrm{Res}_{(0)} \frac{\Delta^2 \mathcal{O}_R \mathcal{O}\prod_{\alpha < \beta} \tilde{m}_{\alpha\beta}}{\prod_{s=1}^{n-1} \prod_{a,b} m^{(s)}_{ab} \prod_\alpha
\det (M_{\alpha})} d\sigma_1 \wedge \cdots \wedge d\sigma_{\mathrm{rk}(G)} = 0
\ee
for any operator $\cO$, where
\[
	\Delta \:=\: \prod_{s=1}^n \prod_{a < b} (\sigma^{(s)}_a - \sigma^{(s)}_b).
\]
Eq. \eqref{localization} implies that $\Delta^2 \prod_{\alpha < \beta} \tilde{m}_{\alpha\beta} \mathcal{O}_R$ is in the ideal generated by $m^{(s)}_{ab}$ and $\det (M_{\alpha})$.
In addition, we have $\Delta\neq 0$ due to the excluded loci condition, $\sigma^{(s)}_a \neq \sigma^{(s)}_b$ if $a \neq b$, and $\prod_{\alpha < \beta} \tilde{m}_{\alpha\beta} \neq 0$ due to $\sigma^{(n)}_a \neq - \sigma^{(n)}_b$ if $a \neq b$, which can be derived from its mirror following \cite{Gu:2018fpm}.

%Let's denote by $x^{(s)}_i$ a generator of degree $i$. All the operators in the quantum cohomology ring can be expressed as polynomials in these generators for $s=1,\cdots,n+1,~i \geq 1$. 
According to the discussion above and the homogeneity of quantum cohomology ring relations, for every operator $\mathcal{O}_R$ in $R=( R_1+R_2 )$, there exists a function $\mathcal{F}_R$ such that
\begin{equation}\label{22relation}
\Delta^2 \prod _{\alpha < \beta } \tilde{m}_{\alpha \beta} \mathcal{O}_R = \mathcal{F}_R({m^{(s)}_{ab}},\det(M_a)).
\end{equation}
Upon imposing the equations of motion \eqref{EOM}, we have
\[
\mathcal{F}_R({m^{(s)}_{ab}},\det(M_a)) = \mathcal{G}_R(q_s,m^{(s)}_{ab},\tilde{m}_{\alpha\beta})
\]
for some function $\mathcal{G}_R$, which can be written as
\[
\mathcal{G}_R = \Delta^2 \prod _{\alpha < \beta } \tilde{m}_{\alpha \beta} \hat{\mathcal{O}}_R(q_s)
\]
for some operator $\hat{\mathcal{O}}_R$ depending on the K\"ahler moduli parameterized by $q_s$.
Consequently, from \eqref{22relation}, we get the quantum ring relation
\begin{equation}\label{relation}
\mathcal{O}_R = \hat{\mathcal{O}}_R(q_s).
\end{equation}

The corresponding classical ring relation is simply
\[
\mathcal{O}_R = 0
\]
since $\hat{\mathcal{O}}_R(0) = 0$. As such, the concrete expression of $\mathcal{F}_R$ can be obtained from the classical ring relations. In order to do so, we just need to take a set of operators generating the classical ring relations, multiply each of them by $\Delta^2 \prod _{\alpha < \beta } \tilde{m}_{\alpha \beta}$ and rewrite the results as functions in ${m^{(s)}_{ab}}$ and $\det(M_a)$. Once the functions $\mathcal{F}_R$ are found, quantum ring relations can be obtained as outlined above.

%In order to write out these classical cohomology rings, the ideals $I$ and $R$ more explicitly, and compare with results in existing literature,

%Let's denote the classical limit of $R$ by $R_0$. 
Now the algorithm computing the quantum cohomology ring can be summarized as the following steps:
\begin{itemize}
    %\item Choose a set of generators of $R = (R_1+R_2)$ in the classical limit.
	\item For every classical relation $\mathcal{O}_R$, find $\mathcal{F}_R$ according to \eqref{22relation}.
	\item Express $\mathcal{O}_R$ in terms of $x^{(s)}_i$, using \eqref{Mblock} and the definition of $x^{(s)}_i$ \eqref{xx}.
	\item Find $\mathcal{G}_R$ and $\hat{\mathcal{O}}_R(q_s)$ from the equations of motion \eqref{EOM} and express $\hat{\mathcal{O}}_R(q_s)$ in terms of $x^{(s)}_i$.
	\item The quantum cohomology ring is then
		\[
			\mathbb{C}[x^{(1)}_{i_1},\cdots,x^{(n+1)}_{i_{n+1}}]/(I+R)
		\]
		with $R$ being generated by operators of the form $\mathcal{O}_R-\hat{\mathcal{O}}_R(q_s)$.
\end{itemize}

\subsection{Ring structure}\label{sec:result}

Now we apply the approach described in the last subsection to compute the quantum cohomology ring of $SF(k_1,k_2,\cdots,k_n,2N)$. 

As mentioned before, the quantum cohomology ring has the form given by \eqref{AIR1R2}. The ideal $I$ is free of the quantum deformation and it is generated by the homogeneous components of 
\be
	[x^{(1)}][x^{(2)}]\cdots[x^{(n+1)}] - 1.
\ee

As metioned above, the classical relations encoded in the ideal $R_1$ consist of $x^{(s)}_l=0$ for $s=1,\cdots,n$ and $l > k_s-k_{s-1}$. The same set of relations also appear in the cohomology of the ordinary flag manifold $F(k_1,k_2,\cdots,k_n,2N)$, which restricts the ranks of $\cS_{s}/\cS_{s-1}$ for $s=1,\cdots,n$. The quantum correction to these relations in the two cases are also identical because they are given by the same set of equations on the Coulomb branch. Hence, the computation for the quantum corrections to $R_1$ is the same as the corresponding computation of \cite{G}, which is also reviewed in the appendix, so here we just list the result. $R_1$ is generated by the ring relations
\be
	x^{(1)}_{\ell}\:=\: 0,\quad \text{for}\ \ell \geq k_1,
\ee
and (see also equation~\eqref{Flagrelation1}),
\be\label{eq:qcrel1}
	x^{(s)}_\ell\:=\:
	\left\{ 
	\begin{array}{ll}
		0,  & k_s - k_{s-1} < \ell < k_s - k_{s-2}, \\
		(-)^{k_{s-1}+k_{s-2}-1} q_{s-1} y^{(s-1)}_{\ell - k_{s}+ k_{s-2}},& \ell \geq k_s-k_{s-2}.
	\end{array} 
	\right. 
\ee

Quantum correction to the ideal $R_2$ is particular to symplectic flag manifolds and requires a more detailed computation. Note that only even components of the quotient bundle $\cQ_n/\cS_n^\vee$ are nontrivial due to $c_r(\cQ_n/\cS_n^\vee) = c_r(\cQ_n \oplus \cQ_n^\vee)$ and classically $c_{2r}(\cQ_n/\cS_n^\vee)=0$ for $r>N-k_n$. These relations get quantum corrections and turn out to generate $R_2$. We have seen from \eqref{R2one} that $R_2$ is classically generated by (We denote by $S(x^2)$ the polynomial $S(x_1^2,x_2^2,\cdots,x_k^2)$ for any symmetric polynomial $S$ in $k$ variables.)
\be
	f_r\:=\: \sum_{i+j=2r}(-)^i h_i(\sigma^{(n)}) h_j(\sigma^{(n)}) \:=\: h_r({\sigma^{(n)}}^2),
\ee
from which we wish to find the expression for $\mathcal{F}_R$. 
Define
\[
	\tilde{\Delta} = \Delta^{(n)} \tilde{m} = \prod_{\alpha < \beta} ({\sigma_\alpha^{(n)}}^2 - {\sigma_\beta^{(n)}}^2),
\]
where
\[
	\tilde{m} = \prod_{\alpha<\beta} (\sigma_\alpha^{(n)} + \sigma_\beta^{(n)}).
\]
We have
\begin{align}
	&\widetilde{\Delta} f_r \:=\: \widetilde{\Delta} h_r({\sigma^{(n)}}^2) \:=\: \sum_{a=1}^{k_n} (-)^a \left(\sigma_a^{(n)} \right)^{2(k_n-1+r)} \tilde{\theta}_a, \no \\
	&=\:(-)^{k_n-1}q_n \left(\prod_{a<b}\left(\sigma_a^{(n)}+\sigma_b^{(n)}\right) \right) \sum_{a=1}^{k_n}(-)^{a-1} \left(\prod_{b=1}^{k_{n-1}}\left(\sigma_b^{(n-1)}-\sigma_a^{(n)}\right) \right) \left(\sigma_a^{(n)}\right)^{2(r+k_n-N-1)} \theta_a,\no \\
	&=\: (-)^{k_{n-1}+k_n-1}q_n \left(\prod_{a<b}\left(\sigma_a^{(n)}+\sigma_b^{(n)}\right) \right) \sum_{a=1}^{k_n}(-)^{a-1} \sum_{i=1}^{k_{n-1}} (-)^i e_{i}\left(\sigma^{(n-1)}\right) \left(\sigma_a^{(n)}\right)^{k_{n-1} - i + 2(r+k_n-N-1)} \theta_a, \no \\
	&=\: \tilde{\Delta} (-)^{k_{n-1}+k_n-1} q_n \sum_{i=1}^{k_{n-1}}(-)^i e_{i}(\sigma^{(n-1)}) h_{2r+ k_{n-1} +k_n -2N - i -1}(\sigma^{(n)}), \no \\
	&=\: \tilde{\Delta} (-)^{k_{n-1}+k_n-1} q_n y_{2r+ k_{n-1} +k_n -2N - 1}^{(n)}. \label{eq:derivation}
\end{align}	
In the above derivation, we have used the following notations for convenience
\begin{align}
	& \theta_a\:=\: \prod_{\substack{i<j\\i,j\neq a}}\left(\sigma^{(n)}_i - \sigma^{(n)}_j  \right),\\
	& \tilde{\theta}_a\:=\: \prod_{\substack{i<j\\i,j\neq a}}\left(\left(\sigma^{(n)}_i\right)^2 - \left(\sigma^{(n)}_j\right)^2  \right),
\end{align}
and for fixed $a$, they satisfy
\be
	\tilde{\theta}_a \prod_{b\neq a} \left(\sigma^{(n)}_a + \sigma^{(n)}_b \right)\:=\: {\theta}_a \prod_{i<j} \left(\sigma^{(n)}_i + \sigma^{(n)}_j \right).
\ee
In the first line of equation~\eqref{eq:derivation}, we have used the Jacobi’s bialternant formula, which says that given a partition $\bm{\lambda} = (\lambda_1,\cdots,\lambda_k)$ with $\lambda_1\geq\lambda_2\geq\cdots\geq\lambda_k\geq 0$ and $\sigma = (\sigma_1,\cdots,\sigma_k)$, we have
\be
	S_{\bm{\lambda}}(\sigma)\:=\: \frac{\det_{i,j}({\sigma_j}^{\lambda_i+k-i})}{\det_{i,j}({\sigma_j}^{k-i})} \:=\: \frac{\det_{i,j}({\sigma_j}^{\lambda_i+k-i})}{\prod_{i<j}({\sigma_i}-\sigma_j)},
\ee
where $S_{\bm{\lambda}}$ is the Schur polynomial of a given partition $\bm{\lambda}$ and the complete symmetric polynomial $h_r(\sigma)$ can be treated as the Schur polynomial of the special partition $\bm{\lambda}=(r,0,\dots,0)$. In the second line of equation~\eqref{eq:derivation}, the last equation of motion in \eqref{EOM} was used and the condition to apply this equation of motion is $2r \geq 2N -k_n - k_{n-1}+1$. Cancelling out the nonzero factor $\widetilde{\Delta}$ in equation~\eqref{eq:derivation} and by definition
\be
	f_r\:=\: \sum_{i+j=2r}(-)^i h_i(\sigma^{(n)})h_j(\sigma^{(n)})\:=\: \sum_{i+j=2r}(-)^i x^{(n+1)}_i x^{(n+1)}_j,
\ee
we have, for $2r \geq 2N -k_n - k_{n-1}+1$,
\be
	\sum_{i+j=2r}(-)^i x^{(n+1)}_i x^{(n+1)}_j \:=\: (-)^{k_{n-1}+k_n-1} q_n y_{2r+ k_{n-1} +k_n -2N - 1}^{(n)}.
\ee
Combined with equation~\eqref{eq:qcrel1}, we obtained the quantum cohomology ring relations as presented in the introduction. 

%********* Here is a comment on the maximal case, namely, when $k_n = N$. In this case, it is known that $\cS^\vee \cong \cQ$, however in the above derivation we have only used the condition that $\cS^\vee$ is a subbundle of $\cQ$. Therefore, in this case we should add another set of classical relations (and there are no quantum corrections to them)
%\be
%	x^{(n+1)}
%\ee
%to make sure $\cS^\vee \cong \cQ$. *********Is this constraint imposed by hand or automatically satisfied?*************

\subsection{Alternative representation in the maximal case}\label{sec:alt}

There exists an alternative representation of the quantum cohomology when $k_n = N$, which will be employed in section \ref{sec:complete}. In this case we can fully take advantage of the property
\be
	\cS_n^\vee \:\cong\: \cQ_n.
\ee
Denote by $g_r$ the right hand side of \eqref{R2two}, then $R_2$ is classically generated by $g_r$, $r \geq 2$ ($g_1 = \sum_{i=1}^{n+1} x^{(i)}_1$ is contained in $I$). For $r \ge 2$, one can compute from \eqref{xx} that
\be\label{eq:gr}
g_r \:=\: \left\{ \begin{array}{ll}
h_r(\sigma^{(n)}) - e_r(\sigma^{(n)}), & 2 \le r \le N, \\
h_r(\sigma^{(n)}), & r > N, \end{array} \right.
\ee
where $h_r$ and $e_r$ are degree-$r$ complete and elementary symmetric polynomials in $\sigma^{(n)}_a$ respectively. In either case, we can rewrite $g_r$ as
\be
	g_r \:=\: \sum_{i=1}^{[r/2]} h_i({\sigma^{(n)}}^2) e_{{r-2i}}(\sigma^{(n)}),
\ee
with conventions
\be
	e_{{0}}(\sigma^{(n)})\:=\: 1,\quad e_{{r}}(\sigma^{(n)})\:=\: 0,\ \text{for}\ r<0,\ \text{or}\ r> k_n=N.
\ee
Thus
\[
\begin{split}
\tilde{\Delta} g_r &= \sum_{m=1}^{[r/2]} \tilde{\Delta} h_m({\sigma^{(n)}}^2) e_{{r-2m}}(\sigma^{(n)})\\
&= \sum_{m=1}^{[r/2]} \left( \sum_{a=1}^{k_n} (-1)^{a-1} (\sigma^{(n)}_a)^{2(k_n-1+m)} \tilde{\theta}_a \right) e_{{r-2m}}(\sigma^{(n)}) \\
&= \sum_{m=1}^{[r/2]} \left( \sum_{a=1}^{k_n} (-1)^{a-1} (\det M_a) (\sigma^{(n)}_a)^{2(m-1)} \tilde{\theta}_a \right) e_{{r-2m}}(\sigma^{(n)}) \\
&= \sum_{m=1}^{[r/2]} \left( \sum_{a=1}^{k_n} (-1)^{a-1} (-q_n \prod_{b=1}^{k_{n-1}} m_{ba}^{(n-1)}) \tilde{m} (\sigma^{(n)}_a)^{2(m-1)} \theta_a \right) e_{{r-2m}}(\sigma^{(n)}) \\
&= (-1)^{k_n+k_{n-1}-1} q_n \tilde{m} \sum_{m=1}^{[r/2]} \left( \sum_{a=1}^{k_n} (-1)^{a-1} \sum_{i=0}^{k_{n-1}} {\sigma^{(n)}_a}^{k_{n-1}-i+2(m-1)} (-1)^i e_{i}(\sigma^{(n-1)}) \theta_a \right) e_{{r-2m}}(\sigma^{(n)}) \\
&= (-1)^{k_n+k_{n-1}-1} q_n \tilde{\Delta} \sum_{m=1}^{[r/2]} \sum_{i=0}^{k_{n-1}}(-1)^i e_{i}(\sigma^{(n-1)}) h_{k_{n-1}-k_n+2m-i-1}(\sigma^{(n)}) e_{{r-2m}}(\sigma^{(n)}),
\end{split}
\]
therefore
\begin{equation}\label{gs}
	g_r \:=\: (-1)^{k_n+k_{n-1}-1} q_n \sum_{m=1}^{[r/2]} (-)^{r-2m} y^{(n)}_{k_{n-1}-k_n+2m-1} y^{(n+1)}_{r-2m}.
\end{equation}
One should keep in mind that $y^{(n+1)}_{r} = 0$ if $r>k_n$.

In the derivation above, we used
\[
	[x^{(s)}] [y^{(s)}] = 1,
\]
\[
	y^{(s)}_r = \sum_{i+j=r} (-1)^i e_{i}(\sigma^{(s-1)}) h_j(\sigma^{(s)}),
\]
and the last identity of equation~\eqref{EOM}.

In summary, quantum ring relations encoded in $R_2$ can be generated by
\be
	g_r\:=\: (-1)^{k_n+k_{n-1}-1} q_n \sum_{m=1}^{[r/2]} (-)^{r-2m} y^{(n)}_{k_{n-1}-k_n+2m-1} y^{(n+1)}_{r-2m},~~r \geq 2.
\ee
This gives an alternative representation of the quantum cohomology in the case $k_n = N$.

\section{Reduction to previous results}\label{sec:reduction}

We have obtained a general representation for arbitrary symplectic flag manifold in terms of generators and generating relations. 
In this section, we focus on two special cases: symplectic Grassmannian and complete symplectic flag manifold. The quantum cohomology rings of both cases have already been found and proved in the literature \cite{KT,BKT,K,GSZ}. We will show that our result reproduces the existing results when specialized to these two cases.

\subsection{Symplectic Grassmannian}

Let us start with the symplectic Grassmannian $SG(k,2n)$ with $k\leq n$. When $k$ is maximal, $k=n$, it is known as the Lagrangian Grassmannian. The equations of motion obtained from the corresponding GLSM are
\be
\label{eq:chiralsg}
	q \prod_{b\neq a}(\sigma_a+\sigma_b)\:=\: \sigma_a^{2n},\quad a,b = 1,\cdots, k,
\ee
which is the special case of equation~\eqref{EOM} with only the last equation being present. 

Now the relation encoded in the ideal $I$ is just 
\be
	[x^{(1)}][x^{(2)}]\:=\:1.
\ee
Comparing with the definition of $y^{(1)}$
\be
	[x^{(1)}][y^{(1)}] \:=\: 1,
\ee
we see that $[x^{(2)}]$ and $[y^{(1)}]$ coincide. As such, we replace $[x^{(2)}]$ with $[y^{(1)}]$ and omit all the superscripts for ease of notation in this subsection.
	
The ideal $R_1$ is generated by
\be
	x_{\ell},\ \ell> k
\ee
and we have seen that there are no quantum corrections to $R_1$. When written in terms of elementary symmetric polynomials in the $\sigma$'s, these relations are easily seen to hold by definition.

For the ideal $R_2$, we can take the result obtained in section \ref{sec:result} and get the generating relations
\be\label{eq:qcsg}
	f_r\:=\: \sum_{i+j=2r}(-)^i h_i(\sigma) h_j(\sigma) \:=\: q h_{2r-2n+k-1}(\sigma),
\ee
or equivalently
\[
\begin{split}
c^2_r &+ 2 \sum_{i=1}^{2N-k-r} (-1)^i c_{r+i} c_{r-i} = (-1)^{2N-k-r} c_{2r+k-2N-1} q,\\
&N-k+1 \leq r \leq N,
\end{split}
\]
where $c_r$ is the $r$-th Chern class of the quotient bundle $\mathcal{Q}$ over $SG(k,2N)$,
which exactly matches the result in \cite{BKT} (notice that $c_r = x^{(2)}_r$). When it comes to the maximal case, $k=n$, we should further apply the constraint $\cQ\cong\cS^\vee$ for $SG(n,2n)$, which implies that $h_r(\sigma) = e_r(\sigma)$. Therefore, we can substitute $h_r$ with $e_r$ in equation~(\ref{eq:qcsg}) and obtain the same results as \cite{KT}.

\subsection{Complete symplectic flag manifold}\label{sec:complete}

Now let us consider the complete symplectic flag manifold, $SF(1,2,\cdots,n,2n)$. This model has already been studied by Kim \cite{K} using Toda lattice model. For comparison with our result, we summarize the result of \cite{K} in below. According to \cite{K}, the quantum cohomology of $SF(1,2,\cdots,n,2n)$ is given by homogeneous components of
\be\label{eq:math}
	\Det\left( \mathbb{1}_{(2n+1)\times(2n+1)} + H_n \right) \:=\: 1.
\ee
(We have set the formal variable $t$ adopted in \cite{K} to one for convenience.) $H_n$ is a $(2n+1)\times(2n+1)$-matrix defined as
\be
	H_n\:=\:
	\begin{bmatrix}
		x^{(1)}_1 & -1 	&0	&0 &0 &0 &0 &0 &0\\
		q_1 &\ddots  &\ddots	&0 &0 &0 &0 &0 &0\\
		0 &\ddots  &\ddots	&-1 &0 &0 &0 &0 &0 \\
		0 &0  &q_{n-1}	&x^{(n)}_1 &0 &0 &0 &0 &-1\\
		0 &0 &0 &0  &-x^{(1)}_1 & -q_1 &0 &0 &0 \\
		0 &0 &0 &0  &1 &\ddots &\ddots &0 &0 \\
		0 &0 &0 &0 &0  &\ddots &\ddots &-q_{n-1} &0\\
		0 &0 &0 &0 &0  &0 &1 &-x^{(n)}_1&-\frac{q_n}{2}\\
		0 &0 &0 &0 &\frac{q_n}{2} &0 &0 &1 &0
	\end{bmatrix}.
\ee
Denote $I_n\equiv \Det\left( \mathbb{1}_{(2n+1)\times(2n+1)} + H_n \right)$ and one observation is that there exists a recursion relation for $I_n$. First, note that the expression of $I_n$ has the following form:
\be
	I_n\:=\: J^+_n J^-_n + \frac{q_n}{2}\left( J^+_n J^-_{n-1} + J^+_{n-1} J^-_n \right),
\ee
with
\be
	J^+_n\:=\: \Det\left(
	\begin{bmatrix}
		1+x^{(1)}_1 & -1 	&0	&0 \\
		q_1 &\ddots  &\ddots &0\\
		0 &\ddots  &\ddots	 &-1 \\
		0 &0  &q_{n-1}	&1+x^{(n)}_1
	\end{bmatrix}\right),
\ee
and
\be
	J^-_n\:=\: \Det\left(
	\begin{bmatrix}
		&1-x^{(1)}_1 & -q_1 &0 &0 \\
		&1 &\ddots &\ddots &0  \\
		&0  &\ddots &\ddots &-q_{n-1}\\
		&0  &0 &1 &1-x^{(n)}_1
	\end{bmatrix} \right).
\ee
As the determinant of a tridiagonal matrix, $J^{\pm}_n$  satisfies the recursion relation
\be
	J^{\pm}_n\:=\: (1\pm x^{(n)}_1) J^{\pm}_{n-1} + q_{n-1} J^{\pm}_{n-2},
\ee
with initial conditions $J^{\pm}_0=1$ and $J^{\pm}_1=1 \pm x^{(1)}_1$.

Now let us show that our result obtained in section \ref{sec:alt} can be reduced to the result of \cite{K}. For convenience, the result of section \ref{sec:alt} for $SF(1,2,\cdots,n,2n)$ can be summarized as follows:
\begin{itemize}
	\item[-] $I$: 
	\be\label{eq:cfI}
		[x^{(1)}][x^{(2)}]\cdots[x^{n}][x^{n+1}] = 1
	\ee
	\item[-] $R_1$:
	\be
		x^{(s)}_\ell\:=\: q_{s-1} y^{(s-1)}_{\ell - 2},\quad \text{for}\ \ell \geq 2. 
	\ee
	In this case, we can further rewrite it as
	\be\label{eq:cfR1}
		[x^{(s)}] \:=\: 1 + x^{(s)}_1 + q_{s-1}[y^{(s-1)}].
	\ee
	\item[-] $R_2$:
	\be
		g_r\:=\: q_n \sum_{m=1}^{[r/2]} (-)^{r-2m} y^{(n)}_{2m-2} y^{(n+1)}_{r-2m}, \quad \text{for}\ r\geq 2,
	\ee
where $g_0=1$ and $g_r$ is the right hand side of \eqref{R2two} for $r>0$.
	As mentioned before, we shall use the convention $y^{(n+1)}_{r-2m} = 0$ if $r-2m>n$ in the summation above. Therefore, we can define
	\be\label{totalG}
		[g]\:\equiv\:\sum_{r=0}^\infty g_r\:=\: \frac{q_n}{2}\left( [y^{(n)}] + [y^{(n)}]^\vee  \right)[y^{(n+1)}]^\vee,
	\ee
	where $[y^{(n)}] = \sum_{r=0}^{\infty} y^{(n)}_r$ and $[y^{(n)}]^\vee = \sum_{r=0}^{\infty} (-)^r y^{(n)}_r$ (similarly for $[y^{(n+1)}]^\vee$). According to equation~\eqref{R2two}, we can write \eqref{totalG} as
	\be\label{eq:cfR2}
		[x^{(n+1)}] - [x^{(1)}]^{\vee}[x^{(2)}]^{\vee}\cdots [x^{(n)}]^{\vee}\:=\: \frac{q_n}{2}\left( [y^{(n)}] + [y^{(n)}]^\vee  \right)[y^{(n+1)}]^\vee.
	\ee
\end{itemize}
The method to show that the above result is the same as equation~\eqref{eq:math} is to apply the relations contained in $R_1$ and $R_2$ to $I$. First note that 
\be\label{Jplus}
	J^+_n\:=\:[x^{(1)}][x^{(2)}]\cdots [x^{(n)}]
\ee
due to $R_1$. This is true because $J^+_1 = [x^{(1)}]$, $J^+_2 = [x^{(1)}][x^{(2)}]$ and 
\be
\begin{split}
	[x^{(1)}][x^{(2)}]\cdots [x^{(n)}] &\:=\: [x^{(1)}][x^{(2)}]\cdots [x^{(n-1)}]\left( 1 + x^{(n)}_1 +  q_{n-1}[y^{(n-1)}]\right)\\
	&\:=\: \left( 1 + x^{(n)}_1 \right) [x^{(1)}][x^{(2)}]\cdots [x^{(n-1)}] + q_{n-1} [x^{(1)}][x^{(2)}]\cdots [x^{(n-2)}]
\end{split}
\ee
shares the same recursion relation with $J^+_n$. The same argument also gives
\be
	J^-_n\:=\:[x^{(1)}]^{\vee}[x^{(2)}]^{\vee}\cdots [x^{(n)}]^{\vee}.
\ee
One can multiply \eqref{eq:cfR2} by \eqref{Jplus} to get
\be\label{eq:equal}
\begin{split}
	[x^{(1)}][x^{(2)}]\cdots[x^{n}][x^{n+1}]&\:=\: J^+_n \left( J^{-}_n + \frac{q_n}{2}\left( [y^{(n)}] + [y^{(n)}]^\vee  \right)[y^{(n+1)}]^\vee \right) \\
	&\:=\: J^+_n J^{-}_n + \frac{q_n}{2} \left( J^+_n J^-_{n-1} + J^+_{n-1} J^-_n \right).
\end{split}
\ee
In the second line, we have used the fact that $[x^{(s)}][y^{(s)}]=1$ and $J^+_n = [x^{(1)}][x^{(2)}]\cdots [x^{(n)}] = [y^{(n+1)}]$. 
%Once obtain equation~\eqref{eq:equal}, we complete the demonstration.
Hence, equation \eqref{eq:cfI} suggests that both sides of \eqref{eq:equal} are equal to one, which exactly reproduces \eqref{eq:math}.

\section{Comments on (0,2) deformation}\label{sec:02}

Quantum sheaf cohomology is a generalization of quantum cohomology \cite{KS}. Its underlying vector space is of the form
\[
H^\bullet (X, \wedge^\bullet \mathcal{E}^\vee)
\]
for some holomorphic vector bundle $\mathcal{E}$ over the K\"ahler manifold $X$. When $\mathcal{E} \cong TX$, quantum sheaf cohomology reduces to quantum cohomology. Quantum sheaf cohomology can be computed by A/2-correlation functions of the corresponding 2d theory with $\mathcal{N}=(0,2)$ supersymmetry and it is topological at least in a neighborhood of the (2,2) locus on the moduli space \cite{ADE} (in this situation the vector bundle $\mathcal{E}$ is a deformed tangent bundle). This deformation will preserve ${\rm ch}_2(TX) = {\rm ch}_2(\mathcal{E})$, namely the anomaly cancellation condition is still satisfied. 

Ring structures of quantum sheaf cohomology have been computed for deformed tangent bundles over toric varieties \cite{MM1, MM2, DGKS}, Grassmannians \cite{GLS} and flag manifolds \cite{G}.
In this section, we attempt to generalize our previous discussions in Section~\ref{sec:computation} by turning on $(0,2)$ deformations and compute the quantum sheaf cohomology for symplectic flag manifolds. However, it turns out that the constraints imposed by supersymmetry forbid any nontrivial $(0,2)$ deformation of the quantum cohomology.

\subsection{(0,2) deformations}
The GLSM describing symplectic flag manifolds was given in section \ref{sec:computation}. Let us turn on the most general deformations of the GLSM and discuss the constraints imposed by supersymmetry and the symplectic form. We start with the symplectic Grassmannians and then generalize the discussion to other symplectic flag manifolds.

\subsubsection*{Symplectic Grassmannian}

The symplectic Grassmannian $SG(k,2n)$ can be described by a GLSM with gauge group $U(k)$, $2n$ chiral fields $\Phi^a_{\pm i}$, $i=1,\cdots,n, a=1,\cdots,k$ in the fundamental representation ${\bf k}$, one chiral field $q_{ab}$ in the representation $\wedge^2 \overline{\bf k}$, and superpotential
\[
W = \sum_{i=1}^n q_{ab} \Phi^a_i \Phi^b_{-i}.
\]
In order to study (0,2) deformation, let's split the chiral fields $\Phi^a_{\pm i}$
into (0,2) chiral fields $\Phi^a_{\pm i}$ and fermi fields $\Psi^a_{\pm i}$, and $q_{ab}$ into (0,2) chiral fields $q_{ab}$ and fermi fields $\Lambda_{ab}$. We only consider linear deformations of the $E$-terms because deformations of $J$-terms and nonlinear deformations of $E$-terms do not change the quantum sheaf cohomology. For the ambient Grassmannian, the deformations of the $E$-terms are encoded in two matrices $A$ and $B$ through \cite{GLS}
\begin{equation}
\bar{D}_+ \Psi^a_{\pm i} = A_i^j \sigma_b^a \Phi^b_{\pm j} + B_{\pm i}^{j} (\mathrm{Tr} \sigma) \Phi^a_{j} + B_{\pm i}^{-j} (\mathrm{Tr} \sigma) \Phi^a_{-j}.
\end{equation}
Similarly, we can deform the $E$-term associated with $q_{ab}$:
\begin{equation}
\bar{D}_+ \Lambda_{ab} = \alpha (-\sigma_a^c q_{cb} - \sigma_b^c q_{ac}) + \beta (\mathrm{Tr} \sigma) q_{ab}.
\end{equation}
In the (2,2) case we have $A=\mathbb{1},B=0,\alpha=1,\beta=0$. By field redefinition, we can set $A=\mathbb{1},\alpha=1$.
To maintain (0,2) supersymmetry, the $E$-terms and $J$-terms must obey
\begin{equation}\label{02constraint}
E^a_i J^i_a + E^a_{-i} J^{-i}_a + \frac{1}{2} E_{ab} J^{ab} = 0,
\end{equation}
where
\[
\begin{split}
&\bar{D}_+ \Psi^a_{\pm i} = E^a_{\pm i},\quad \bar{D}_+ \Lambda_{ab} = E_{ab} \\
&J^{\pm i}_a = \frac{\partial W}{\partial \Phi^a_{\pm i}} = \pm q_{ab} \Phi^b_{\mp i}, \\
&J^{ab} = \frac{\partial W}{\partial q_{ab}} = \sum_{i=1}^n \left( \Phi^a_i \Phi^b_{-i} - \Phi^b_i \Phi^a_{-i} \right).
\end{split}
\]
One can compute that
\[
\begin{split}
&E^a_i J^i_a + E^a_{-i} J^{-i}_a + \frac{1}{2} E_{ab} J^{ab} \\
= &(\mathrm{Tr} \sigma) q_{ab} (\phi_i^b,\phi_{-i}^b) \left(
\begin{array}{cc} 0 & -\mathbb{1} \\ \mathbb{1} & 0 \end{array} \right) \left[ \left(
\begin{array}{cc} B_i^j & B_i^{-j} \\ B_{-i}^j & B_{-i}^{-j} \end{array} \right)
\left( \begin{array}{c} \phi^a_j \\ \phi^a_{-j} \end{array}\right) + \frac{1}{2} \beta \left( \begin{array}{c} \phi^a_i \\ \phi^a_{-i} \end{array}\right) \right],
\end{split}
\]
which implies
\[
B = -\frac{\beta}{2} \mathbb{1}_{2n \times 2n}
\]
by \eqref{02constraint}. If $A$ and $\alpha$ are not fixed, then \eqref{02constraint} would imply
\[
A = \alpha \mathbb{1}_{2n \times 2n}.
\]
The deformation induced by $\alpha$ can be absorbed into a re-scaling of $\sigma$.

%\subsubsection*{Flag manifolds}

\subsubsection*{Symplectic flag manifold}

%Combining previous two discussions immediately give results for 
In the case of $SF(k_1,k_2,\cdots,k_n,2N)$, we have $P_{ab}$ transforming in $\wedge^2 \overline{\bf k_n}$ and a superpotential
\[
W = \sum_{i=1}^N P_{ab} {\Phi_{(n,n+1)}}^a_i {\Phi_{(n,n+1)}}^b_{-i}.
\]
Let $\Lambda_{s,s+1}$ be the companion fermi multiplet of $\Phi_{s,s+1}$, then, as ordinary flag manifolds reviewed in appendix \ref{sec:app}, we have (0,2) deformations given by
\[
\begin{split}
	&\overline{D}_+ \Lambda_{s,s+1}= \Phi_{s,s+1} \Sigma^{(s)} - \Sigma^{(s+1)} \Phi_{s,s+1} + \sum_{t=1}^n u^s_t ({\rm Tr}\Sigma^{(t)}) \Phi_{s,s+1},\\
	&\quad s=1,\cdots,n-1\\
	&\overline{D}_+ \Lambda_{n,n+1}^{\pm i} = \Phi_{n,n+1} \Sigma^{(n)} + \sum_{t=1}^n({\rm Tr}\Sigma^{(t)}) {(A_t)}_j^{\pm i} \Phi_{n,n+1}^j + \sum_{t=1}^n({\rm Tr}\Sigma^{(t)}) {(A_t)}_{-j}^{\pm i} \Phi_{n,n+1}^{-j},\\
   &\quad i=1,\cdots,N,
\end{split}
\]
where $\Sigma_s$ is the chiral multiplet in the adjoint representation of $U(k_s)$. The parameters $u^s_t$ are constants and $A_t$ are $2N \times 2N$ matrices.
In addition, let $\Lambda_{ab}$ be the companion fermi multiplet of $P_{ab}$, then we have 
\begin{equation}
\bar{D}_+ \Lambda_{ab} = -{\sigma^{(n)}}_a^c P_{cb} - {\sigma^{(n)}}_b^c P_{ac} + \sum_{t=1}^n \beta_t (\mathrm{Tr} \sigma^{(t)}) P_{ab}.
\end{equation}
The linear (0,2) deformations are parameterized by $u^s_t, A_t$ and $\beta_t$, $\mathcal{N}=(2,2)$ supersymmetry is recovered if they all vanish.
Since
\[
J^{ab}_{(s,s+1)}= \frac{\partial W}{\partial \Phi_{(s,s+1)}^{ab}} = 0,
\]
for $1 \leq s \leq n-1$,
$\mathcal{N}=(0,2)$ supersymmetry requires
\[
{J_{(n,n+1)}}^i_a {E_{(n,n+1)}}^a_i+{J_{(n,n+1)}}^{-i}_a {E_{(n,n+1)}}_{-i}^a+ \frac{1}{2} \tilde{J}^{ab} \tilde{E}_{ab} = 0,
\]
where
\[
{J_{(n,n+1)}}^i_a = \frac{\partial W}{\partial {\Phi_{(n,n+1)}}^a_i},
\]
\[
\tilde{J}^{ab} = \frac{\partial W}{\partial q_{ab}}.
\]
One can show as in the case of symplectic Grassmannian that
\[
A_t = -\frac{1}{2} \beta_t \mathbb{1}.
\]
If we define $u^n_t \equiv -\frac{1}{2} \beta_t$, then the (0,2) deformation is parameterized by the parameters $u^s_t, s,t=1,2,\cdots,n$.

\subsection{Quantum sheaf cohomology}
So far we have derived the possible $(0,2)$ deformations, in this subsection we attempt to extract the quantum sheaf cohomology. Again we start with the Coulomb branch of the corresponding $(0,2)$ model. On the Coulomb branch, the equations of motion take the same form as equation~\eqref{EOM} but the matrices $m^{(a)}$, $\tilde{m}$ and $M_{\alpha}$ are deformed as follows
\be\label{eq:Mblockdeformed}
\begin{split}
	m^{(s)}_{a b} &= \sigma^{(s)}_a-\sigma^{(s+1)}_b + \sum_{t=1}^n u_t^s \mathrm{Tr}(\sigma^{(t)}),\\
 	& \quad s=1,\cdots, n-1,~~a=1,\cdots,k_{s},~~b=1,\cdots,k_{s+1}, \\
	M_a &= \sigma^{(n)}_a \mathbb{1}_{2N\times2N} + \sum_{t=1}^n (\mathrm{Tr}\sigma^{(t)}) u^n_t \mathbb{1}_{2N\times2N}, \\
	\tilde{m}_{ab} &= -\sigma^{(n)}_a-\sigma^{(n)}_b-2 \sum_{t=1}^n u^n_t (\mathrm{Tr}\sigma^{(t)}), \quad a \neq b.
\end{split}
\ee
For simplicity, let $h_s =\sum_{t=1}^n u_t^s \mathrm{Tr}(\sigma^{(t)}) $ for $s= 1,\cdots,n$. Then if we redefine
\begin{align*}
	& \widetilde{\sigma}^{(s)}_a\:=\: \sigma^{(s)}_a + \sum_{\ell=s}^n h_l,
\end{align*}
equation~\eqref{eq:Mblockdeformed} has the exact same form as equation~\eqref{Mblock}:
\be
\begin{split}
	m^{(s)}_{a b} &= \widetilde{\sigma}^{(s)}_a - \widetilde{\sigma}^{(s+1)}_b,\\
	& \quad s=1,\cdots, n-1,~~a=1,\cdots,k_{s},~~b=1,\cdots,k_{s+1}, \\
	M_a &= \widetilde{\sigma}^{(n)}_a \mathbb{1}_{2N\times2N}, \\
	\tilde{m}_{ab} &= -\widetilde{\sigma}^{(n)}_a-\widetilde{\sigma}^{(n)}_b, \quad a \neq b.
\end{split}
\ee
This shows that the equations of motion will induce the same gauge invariant relations as those obtained without $(0,2)$ deformation as long as the generators are interpreted as symmetric polynomials in the $\tilde{\sigma}$'s instead of the $\sigma$'s.
Put another way, the $(0,2)$ deformations presented in the generating relations can all be absorbed into redefinition of the generators. Thus the quantum cohomology ring remains the same. It is in this sense we say that there are no nontrivial $(0,2)$ deformations\footnote{If the (0,2) deformation is diagonal, one can also perform the same derivation through the mirror theory proposed in \cite{Gu:2019byn}.}.

\section{Summary and outlook}

In this paper, we derived the quantum cohomology ring of arbitrary symplectic flag manifolds from the non-abelian GLSM construction, generalizing the previous results. In two special cases of the symplectic flag manifolds, namely the symplectic Grassmannians and the complete symplectic flag manifolds, our results agree with \cite{KT,BKT,K}. 

By performing $(0,2)$ deformations of the GLSM, we found that there are no nontrivial deformations to the quantum cohomology ring. In principle, one can also consider other deformations, such as the twisted mass deformation. These twisted masses are holonomies associated with the global symmetry. Because the superpotential restricts the global symmetry and the number of independent twisted masses is equal to the rank of the global symmetry group, the possible twisted mass deformations are restricted by the form of the superpotential (\ref{eq:superpotential}). With this deformation, one will obtain the equivariant quantum cohomology for symplectic flag manifolds. Our algorithm in this paper can be generalized to the equivariant case straightforwardly.

There are several directions worth exploring along this line. The localization formula in \cite{CCP} is more general than the one we employed in this paper. In particular, it includes an equivariant deformation of the two-sphere (the A-twist corresponds to zero deformation). The correlators on the deformed two-sphere lead to deformation of the quantum cohomology, whose meaning and structure deserve further study. On the other hand, the authors of \cite{Gu:2020zpg} discussed how to extract the quantum K-theory ring relations from the corresponding 3d Chern-Simons-matter theory, which is a $S^1$-lift of an $\cN=(2,2)$ GLSM. It is natural to consider the $S^1$-lift of an $\cN=(0,2)$ GLSM and the K-theoretic extension of quantum sheaf cohomology.

\section*{Acknowledgement}

We would like to thank Wei Gu, Leonardo Mihalcea, Mauricio Romo and Eric Sharpe for comments and suggestions. We thank in particular Eric Sharpe for collaboration at the beginning of this project. JG acknowledges support from the China Postdoctoral Science Foundation No. 2020T130353.

\appendix

\section{Quantum sheaf cohomology of flag manifolds}\label{sec:app}

In this appendix we review the derivation of \cite{G} for the quantum sheaf cohomology of ordinary flag manifolds\footnote{The relationship between the vertex partition function and the $I$-function describing the equivariant quantum cohomology of flag manifolds can be found in \cite{BSTV}.}. We will obtain an equivalent but simpler form of the ring structure by using a different set of generators.

For every sequence of integers $(k_1,k_2,\cdots,k_n)$ with $0 < k_1 < k_2 < \cdots < k_n < N$, the flag manifold $F(k_1,k_2,\cdots,k_n,N)$ is defined by the set of flags in $\mathbb{C}^N$:
\[
F(k_1,k_2,\cdots,k_n,N) = \{(V_{k_1},\cdots,V_{k_n}) \in G(k_1,N) \times \cdots \times G(k_n,N) | V_{k_1} \subset \cdots \subset V_{k_n}\}.
\]
The $\cN=(2,2)$ GLSM describing $F(k_1,k_2,\cdots,k_n,N)$ is a quiver gauge theory with gauge group $U(k_1) \times \cdots \times U(k_n)$ \cite{DS}. For each $s=1,\cdots,n-1$, there is a chiral multiplet $\Phi_{s,s+1}$ transforming in the fundamental representation of $U(k_s)$ and in the antifundamental representation of $U(k_{s+1})$. There are also chiral multiplets $\Phi_{n,n+1}^i$ transforming in the fundamental representation of $U(k_n)$ for $i=1,\cdots,N$.

There is a flag of universal subbundles
\[
	0=\mathcal{S}_0 \hookrightarrow \mathcal{S}_1 \hookrightarrow \mathcal{S}_2 \hookrightarrow \cdots \hookrightarrow \mathcal{S}_n \hookrightarrow \mathcal{S}_{n+1}=\mathcal{O}^{\oplus N},
\]
where the fibers of these bundles at any point of the flag manifold form the flag corresponding to that point, so $\mathcal{S}_i$ has rank $k_i$.

The deformed tangent bundle of $F(k_1,k_2,\cdots,k_n,N)$ can be described by turning on (0,2) deformations of the quiver GLSM. These deformations are encoded in the $E$-terms. Denote by $\Lambda_{i,i+1}$ the fermi multiplet corresponding to the chiral multiplet $\Phi_{i,i+1}$, i.e. $\Lambda_{i,i+1}$ and $\Phi_{i,i+1}$ combine into an $\cN=(2,2)$ chiral multiplet when the deformations are turned off. Up to field redefinitions, the $E$-terms with the most general linear deformations are given by\footnote{Nonlinear deformations do not change A/2 correlation functions \cite{CGJS} and as such they do not affect the quantum sheaf cohomology ring. }
\begin{equation}\label{E_F}
\begin{split}
	&\overline{D}_+ \Lambda_{s,s+1}= \Phi_{s,s+1} \Sigma^{(s)} - \Sigma^{(s+1)} \Phi_{s,s+1} + \sum_{t=1}^n u^s_t ({\rm Tr}\Sigma^{(t)}) \Phi_{s,s+1},\\
	&\quad s=1,\cdots,n-1\\
	&\overline{D}_+ \Lambda_{n,n+1}^i= \Phi_{n,n+1} \Sigma^{(n)} + \sum_{t=1}^n({\rm Tr}\Sigma^{(t)}) {A_t}_j^i \Phi_{n,n+1}^j, ~i=1,\cdots,N,
\end{split}
\end{equation}
where we have suppressed the gauge indices. $\Sigma_s$ is the chiral multiplet in the adjoint representation of $U(k_s)$. The parameters $u^s_t$ are constants and $A_t$ are $N \times N$ matrices. When $A_t=u^s_t=0$, we recover the $\cN=(2,2)$ GLSM for the flag manifold.

At a generic point of the Coulomb branch, the gauge group is spontaneously broken to $\prod_{s=1}^{n} U(1)^{k_s}$. The massless degrees of freedom are the diagonal entries of $\sigma^{(s)}, s=1,\cdots,n$. We denote by $\sigma^{(s)}_a$ the $a$-th diagonal element of $\sigma^{(s)}$ and $\mathrm{Tr}\sigma^{(s)} = \sum_{a=1}^{k_s} \sigma^{(s)}_a$. The mass matrices are
\begin{equation}\label{mM}
\begin{split}
	m^{(s)}_{a b} &= \sigma^{(s)}_a-\sigma^{(s+1)}_b + \sum_{t=1}^n u_t^s \mathrm{Tr}(\sigma^{(t)}),\\ 
	s&=1,\cdots, n-1,~~a=1,\cdots,k_{s},~~b=1,\cdots,k_{s+1}, \\
	M_a &= \sigma^{(n)}_a I + \sum_{t=1}^n (\mathrm{Tr}\sigma^{(t)}) A_t.
\end{split}
\end{equation}

The method used in \cite{G} is essentially the same as the one we used in this paper, but there is one difference. In the GLSM describing ordinary flag manifolds, the field $P_{ab}$ in $\wedge^2\overline{\bf k_n}$ is absent. As a consequence, the localization formula suggests that $\Delta^2 \mathcal{O}_R$ is in the ideal generated by $m^{(s)}_{ab}$ and $\det M_a$ ($\tilde{m}_{ab}$ is gone) for any classical ring relation $\mathcal{O}_R$.

On the (2,2) locus, we interpret $\sigma^{(s)}_a$ as the Chern roots of $\mathcal{S}_s^\vee$. Let $x^{(s)}_r$ be the $r$-th Chern class of $\mathcal{S}_s/\mathcal{S}_{s-1}$, then we can write $x^{(s)}_r$ as polynomials in $\sigma$:
\begin{equation}\label{x}
\begin{split}
	x^{(1)}_r &= (-1)^r e_r(\sigma^{(1)}), \\
	x^{(s)}_r &= \sum_{i+j=r} h_i(\sigma^{(s-1)})(-1)^j e_j(\sigma^{(s)}),~s=2,\cdots, n, \\
	x^{(n+1)}_r &= h_r(\sigma^{(n)}),
\end{split}
\end{equation}
where $h_r$ and $e_r$ are the degree-$r$ complete and elementary symmetric polynomials repectively.
For fixed $s \leq n$, the degree $r$ component of $[x^{(1)}][x^{(2)}] \cdots [x^{(s)}]$ is $(-1)^r e_r(\sigma^{(s)})$. Consequently, all the polynomials invariant under the permutation of $(\sigma^{(s)}_1,\cdots,\sigma^{(s)}_{k_s})$ for all $s$ can be generated by $x^{(i)}, i=1,\cdots,n+1$.

Note that $[x^{(s)}][y^{(s)}]=1$ implies
\begin{equation}\label{y}
\begin{split}
	y^{(1)}_r &= h_r(\sigma^{(1)}), \\
	y^{(s)}_r &= \sum_{i+j=r} (-1)^i e_i(\sigma^{(s-1)}) h_j(\sigma^{(s)}),~s=2,\cdots, n, \\
	y^{(n+1)}_r &= (-1)^r e_r(\sigma^{(n)}).
\end{split}
\end{equation}

The classical cohomology ring of $F(k_1,k_2,\cdots,k_n,N)$ can be written as
\[
\mathcal{A}/( I+R ),
\]
where $\mathcal{A}$ is the polynomial ring generated by the $x^{(s)}_r$'s. $I$ encodes
\[
[x^{(1)}][x^{(2)}]\cdots[x^{(n+1)}] = 1.
\]
$R$ is generated by $x^{(s)}_r$ for $r > k_s - k_{s-1}$, $s=1,\cdots,n+1$, on the (2,2) locus.

Now we compute (0,2) deformation and quantum correction to the cohomology ring. For $s=1,\cdots,n-1$, define
\[
	w_s = \sum_{t=1}^n u^s_t \mathrm{Tr}(\sigma^{(t)}),
\]
and
\begin{equation}\label{redef}
	c^{(s)}_a = \sigma^{(s)}_a + \sum_{i=s}^{n-1} w_i,
\end{equation}
then
\[
	m^{(s)}_{ab} = \sigma^{(s)}_a - \sigma^{(s+1)}_b + w_s = c^{(s)}_a - c^{(s+1)}_b.
\]
We see that the deformations parameterized by $u^s_t$ are absorbed by the redefinition \eqref{redef}, therefore they are spurious (0,2) deformations. From \eqref{mM}, we also have
\[
	M_a = \sigma^{(n)}_a I + \sum_{t=1}^n (\mathrm{Tr} \sigma^{(t)}) A_t,
\]
where $I$ is the $N \times N$ identity matrix. The equations of motion derived from the effective $J$-terms on the Coulomb branch read
\begin{equation}\label{EOMFlag}
\begin{split}
	&\prod_{a=1}^{k_2} m^{(1)}_{\alpha a} = (-)^{k_1-1}q_1,~ \alpha=1,\cdots,k_1, \\
	&\prod_{b=1}^{k_{s+1}} m^{(s)}_{\alpha b} = (-)^{k_s-1}q_s \prod_{a=1}^{k_{s-1}} m^{(s-1)}_{a \alpha},~ \alpha=1,\cdots,k_s,~s=2,\cdots,n-1, \\
	&\det (M_\alpha) = (-)^{k_{n}-1}q_n \prod_{a=1}^{k_{n-1}} m^{(n-1)}_{a \alpha},~\alpha=1,\cdots,k_n.
\end{split}
\end{equation}
Since
\[
	\Delta^{(s)} = \prod_{a<b} (\sigma^{(s)}_a - \sigma^{(s)}_b) = \prod_{a<b} (c^{(s)}_a - c^{(s)}_b)
\]
for $s=2,\cdots,n$, we have a set of relations identical to those on the (2,2) locus, namely
\begin{equation}\label{Flagrelation1}
x^{(s)}_l =
\left\{ \begin{array}{ll}
(-)^{k_s+k_{s-2}-1}q_{s-1} y^{(s-1)}_{l-k_s+k_{s-2}}, & l \geq k_s-k_{s-2}, \\
0, & k_s - k_{s-1} < l < k_s - k_{s-2},
\end{array} \right.
\end{equation}
for $s=2,\cdots,n$. These can be derived from the identity
\[
\Delta^{(s)} x^{(s+1)}_{k_{s+1}-k_s+r} = \sum_{a=1}^{k_s} (-1)^{a-1} {c^{(s)}_a}^{r-1} \prod_{\alpha<\beta,\alpha,\beta \neq a} (c^{(s)}_\alpha-c^{(s)}_\beta) \left( \prod_{b=1}^{k_{s+1}} m^{(s)}_{ab} \right),
\]
and the equation of motion for $m^{(s)}_{ab}$ \eqref{EOMFlag}.
Moreover, because $x^{(n+1)}_r = h_r(\sigma^{(n)})$ and
\[
\det M_a = \sum_{i=0}^N {\sigma_a^{(n)}}^{N-i} I_i \left(\sum_{t=1}^n (\mathrm{Tr} \sigma^{(t)}) A_t \right),
\]
where $I_i(\cdot)$ denotes the $i$-th characteristic polynomial, we have
\[
%\begin{split}
\Delta^{(n)} x^{(n+1)}_r = \sum_{a=1}^{k_n}(-1)^a {\sigma_a^{(n)}}^{k_n-N+r-1} \prod_{b<c,b,c\neq a} (\sigma^{(n)}_b - \sigma^{(n)}_c) (\det M_a),  \]
which becomes
\[
%&\stackrel{(0,2)}{\longrightarrow}
\Delta^{(n)} \sum_{i=0}^N  I_i \left(\sum_{t=1}^n (\mathrm{Tr} \sigma^{(t)}) A_t \right) h_{r-i}(\sigma^{(n)})
%\end{split}
\]
off the (2,2) locus.
On the other hand, the equation of motion for $\det M_a$ \eqref{EOMFlag} yields
\[
\sum_{a=1}^{k_n}(-1)^a {\sigma_a^{(n)}}^{k_n-N+r-1} \prod_{b<c,b,c\neq a} (\sigma^{(n)}_b - \sigma^{(n)}_c) (\det M_a) = -q_n (-1)^{k_{n-1}} \Delta^{(n)} y^{(n)}_{r+k_{n-1}-N}.
\]
Thus, if we define
\[
R_r \equiv \sum_{i=0}^{\min\{N,r\}} I_i \left(\sum_{t=1}^n (\mathrm{Tr} \sigma^{(t)}) A_t \right) h_{r-i}(\sigma^{(n)}),
\]
then we have the ring relations
\begin{equation}\label{Flagrelation2}
R_r =
\left\{ \begin{array}{ll}
0, & N-k_n < r < N-k_{n-1}, \\
-q_n (-1)^{k_{n-1}} y^{(n)}_{r+k_{n-1}-N}, & r \geq N-k_{n-1}.
\end{array} \right.
\end{equation}
Note that from \eqref{x}, $\mathrm{Tr}(\sigma^{(t)}) = -(x^{(1)}_1 + \cdots + x^{(t)}_1)$ for $t = 1,\cdots,n$ and $h_{r-i}(\sigma^{(n)}) = x^{(n+1)}_{r-i}$.

In summary, the quantum sheaf cohomology ring of the flag manifold $F(k_1,\cdots,k_n,N)$ is
generated by $x^{(1)}_1$, $x^{(s)}_{i_s}$, $s=2,\cdots,n+1$, $i_s \geq 1$, subject to the ring relations
\eqref{Flagrelation1}, \eqref{Flagrelation2} and $[x^{(1)}] [x^{(2)}] \cdots [x^{(n+1)}] = 1$.

%\bibliography{QCref.bib}
%\bibliographystyle{fullsort}

\end{document}